\documentstyle[multicol,aps,prl,epsf,psfig]{revtex}

\begin{document}  

\title{Phase transitions in the steady state behavior of mechanically
       perturbed spin glasses and ferromagnets}

\author{Alexandre Lef\`evre and David S. Dean\\
{IRSAMC, Laboratoire de Physique Quantique, Universit\'e Paul Sabatier, 118
route de Narbonne, 31062 Toulouse Cedex 04, France}
}
\date{November 13  2001}
\maketitle
 
\begin{abstract}
We analyze the steady state regime of systems interpolating between spin
glasses and ferromagnets under a tapping dynamics
recently introduced by analogy with the 
dynamics of mechanically perturbed granular media. A crossover from a
second order to first order ferromagnetic transition
as a function of the spin coupling distribution is found. The 
flat measure over blocked states introduced by Edwards for granular
media is used to explain this scenario. Annealed calculations
of the Edwards entropy are shown to qualitatively explain the nature of the
phase transitions. A Monte-Carlo construction of the Edwards measure
confirms that this explanation is also quantitatively accurate.

\vskip 0.5cm

\end{abstract}

\begin{multicols}{2}

In complex systems such as granular media the energy available due to thermal
fluctuations is not sufficient to cause particle rearrangement, hence in the
absence of external perturbations the system is trapped in a metastable state.
A granular media may be shaken mechanically and experiments reveal a 
steady state regime defined by an asymptotic density \cite{exps}. 
The non trivial behavior of these systems, such as slow relaxation dynamics
and hysteresis effects arises from the fact that such systems have an
extensive entropy of metastable or blocked states. A vertically
tapped system of hard spheres tends to a random close packing
\cite{exps}, whereas
a horizontally shaken system crystallizes \cite{horiz}, the
stationary states obtained in these systems are not theoretically 
understood for the moment.
Edwards \cite{edwme,phsep} has proposed 
that one may construct 
a thermodynamics over metastable states in the same way as Boltzmann and 
Gibbs  developed over microstates, his hypothesis is that the 
equilibrium measure is
flat over all blocked or metastable states satisfying the relevant 
macroscopic constraints. This is an ergodic hypothesis which is conceivably 
true in the case of {\em extensive manipulations} such as stirring, pouring,
shaking an tapping. This scenario had been recently tested on a wide range
of models by comparing dynamical time averaged quantities in the
steady state with those predicted by a flat measure which is either
calculated numerically or analytically \cite{bar,chmes,oned}. In a number 
of systems the Edwards measure was seen to work extremely well, although,
reassuringly given the strength of the hypothesis,
there are cases where it does not appear to be applicable \cite{bar}.

Given the recent interest in the Edwards measure and the fundamental
and industrial importance of mechanically driven systems it is
natural to examine  phase transitions in this context. As pointed out
by Edwards and coworkers, the construction of a steady state measure
for mechanically perturbed systems would allow one to predict, for
example, whether two powders or granular media form a mixture or phase
separate upon stirring or shaking. In mean field implementations of
the Edwards measure to binary granular systems second order
phase transitions have been found between the mixed and  
separated phases \cite{phsep}.
  
Recently \cite{chmes} there has been interest in using spin systems to examine 
tapping like dynamics on systems with an extensive entropy of metastable
states. These systems, though far from being realistic models of granular
media, provide a testing ground for the development of thermodynamic like
theories for granular systems and indeed seem to show much of the
phenomenology seen in real systems.

The models we shall consider are Ising spin systems on random thin graphs.
A random thin graph is a collection of $N$ points, each point being linked to
exactly $c$ of its neighbors, $c$ therefore 
being the connectivity of the graph.
The spin glass/ferromagnet model we shall consider has the Hamiltonian 
$H = -{1\over 2} \sum_{j\neq i} J_{ij} n_{ij} S_i S_j$
where the $S_i$ are Ising spins, $n_{ij}$ is equal to one if the 
sites $i$ and $j$ are connected and zero otherwise. 
The fact that the local connectivity is fixed as $c$ imposes the
local constraints $\sum_{j} n_{ij} = c$, for all sites $i$. For the
following, we have fixed $c=3$.
The couplings $J_{ij}$ are independent random variables with distribution:
$P(J_{ij})=\alpha \delta(J_{ij}+1)+(1-\alpha) \delta(J_{ij}-1)$ ($\alpha$
being a parameter in $[0,1/2]$). 

We define the tapping dynamics as follows: 
In between the taps, the system has a natural zero
temperature relaxational dynamics, it
evolves under a random sequential single spin flip 
dynamics where only moves which reduce the energy are allowed.
When blocked, it is tapped  with strength  $p\in [0,1/2]$, 
that is to say each spin is flipped with a probability $p$, the updating at 
this point being parallel - this corresponds to
the extensive manipulation. The system is
then evolved by the zero temperature dynamics until it becomes 
once again stuck and the tapping is repeated. After a transient regime,
the system reaches a stationary regime characterized by an energy per spin 
$E(p)$. Numerical simulations show that $E(p)$ is a monotonically
decreasing function of $p$ - the lighter one taps, the lower the energy
obtained. Here we make the analogy with experiments on vertically tapped
granular media \cite{exps}, 
where (in the reversible part of the experiment) 
the asymptotic compactivity increases as the system is more lightly 
tapped. Our numerical simulations are reversible and one can move
up and down the curve of $E(p)$ if the system is tapped for a
sufficiently long time after changing $p$. Here the macroscopic
quantity which is fixed on average by the tapping dynamics is $E(p)$, 
whereas in granular material it is the average volume $v = V/N$ per particle. 

The canonical form of the Edwards hypothesis
\cite{oned,spins} then leads to the consideration of the partition function
$Z(\beta)
=\int\,dE\,dm\,\exp\left(-N(\beta E - s_{Edw}(E,m))\right)$,
where $s_{Edw}(E,m)$ the  Edwards
entropy per spin at average energy $E$ and magnetization
$m$ per spin. Here $\beta = \beta(p)$ is a Lagrange multiplier
fixing $E$, numerical tapping simulations thus suggest that 
$\beta(p)$ decreases 
as $p$ is increased. 
In spin glasses the Edwards entropy per spin 
at fixed average energy can be calculated (or approximated) and can be shown
to be extensive ({\em i.e.} $s_{Edw}(E)$ is of order $1$ and independent
of $N$ in the thermodynamic limit). As mentioned previously granular systems
are expected to share this property. Let us emphasize at this point that
this rather ambitious goal of developing a thermodynamics for granular
systems does use additional information about the dynamics as compared to
usual thermodynamics. The definition of the entropy contains the
information about  metastability. This is a partial, though rather limited,
use of dynamics to describe statics.

If we assume that $\beta(p)$ is continuous on reducing
$p$ one expects the possibility of an onset of ferromagnetic ordering, which
would correspond to phase separation in a granular material. The Edwards
free energy per spin $ f(\beta) = {\rm min}_{E,m} f(E,m,\beta) 
= {\rm min}_{E} f(E,\beta)$ with
$f(E,m,\beta) = E - s_{Edw}(E,m)/\beta$ and $f(E,\beta) 
= f(E,m(E), \beta)$, where 
$m(E)$ is the point where $f(E,m, \beta(p))$ is a minimum with respect to $m$.
The value of $m(E)$ is thus given by  $\partial s_{Edw}(E,m(E))/\partial m =0$.
In the models studied here we find that there exists a value $E_c$ such that
for $E > E_c$ one has $m(E) = 0$ and for $E < E_c$ the solution $m=0$
becomes
a local maximum of $f(E,m,\beta)$ and there is a second order phase transition
with $m$ becoming non zero. 

For $E< E_c$ in the neighborhood of $E_c$
we have found that, depending on the parameters of the
disorder, there are two possible behaviors:

(1) ${\partial^2 s_{Edw}(E,m(E))/\partial E^2}|_{E=E_c^-} <0 $ in which case 
the ferromagnetic phase is stable and the system exhibits a second
order phase transition in the energy with a discontinuity in 
$\partial \beta/\partial E$.

(2) ${\partial^2 s_{Edw}(E,m(E))/\partial E^2}|_{E=E_c^-}  >0 $
and the ferromagnetic phase is unstable. In this case there is 
a second minimum of $f(E)$ (case (2a)) or end point of $f(E)$ (where the 
Edwards entropy vanishes, that is to say the ground state $E_{GS}$)
(case (2b))  at an energy $E^*$ which is strictly
lower than $E_c$. Hence there is a first order phase transition with
a non zero spontaneous
magnetization $m(E^*)$. We find in this case that the paramagnetic phase
may be in fact only metastable for a range of values $\beta < \beta_c$
{\em i.e.} the second minimum exists below $\beta_c$ 
and may have a lower Edwards free energy. Comparing our simulations with
the annealed calculation we find it is only when this metastable phase 
becomes unstable (at the spinodal point) that the transition occurs. 
This suggests that the tapping dynamics is not efficient at tunneling
over free energy barriers - it is interesting to note that the fluffy 
(arch rich) phase in granular media exhibit this strong metastability 
\cite{exps}.

With respect to the relaxational part of the dynamics introduced above, 
the total number of
metastable states for a given set of $J_{ij}$ and $n_{ij}$
is given by : 
\begin{eqnarray}
N_{MS}(E,m)  = {\rm Tr} \prod_{i=1}^N \theta\left( \sum_{j\neq i}
J_{ij}n_{ij} S_i S_j \right)\\ \nonumber
\delta(H-NE)\, \delta(\sum_i S_i - Nm) 
\end{eqnarray}
where $\theta(x) = 0$ if $x< 0$ and $\theta(x) = 1$ if $x\ge 0$. The above 
formula expresses the fact that each spin is aligned with its local molecular
field in a metastable state and hence none can flip under the relaxational
dynamics.
 The Edwards entropy of the metastable states of energy per
spin $E$ and at magnetization $m$ is thus $s_{Edw}(E,m)={1 \over N}\langle\langle 
\ln{N_{MS}}(E,m)\rangle\rangle$, the brackets
denoting the average over the disorder. Unfortunately this 
average is not easily taken  analytically at low energy, and
we have to compute an annealed average, which will give an upper
bound for the Edwards entropy: 
\begin{equation}\label{eqnedw}
s_{Edw}(E,m)=\lim_{N\rightarrow \infty}{1\over N}\ln\langle\langle
N_{MS}(E,m)\rangle\rangle 
\end{equation}
In the the SK model\cite{skms} and for Ising spin systems on 
random thin graphs 
\cite{ms},  there exists an energy $E_{ann}$ above which this  
annealed approximation is exact.

The computation of this 
quantity is a generalization of that for
$\alpha=0$ \cite{ms}. One finds $s_{Edw}(E,m)$ is given 
by the saddle point value of the action
\begin{eqnarray}\label{eqnsm}
&S(u,v,t,\beta;E,m)=-\beta\,(\frac{c}{2}-E)+{c\over 2}\ln(1-2\alpha)\\ 
\nonumber
              &-{c\over 2}\ln
\left[(1-\alpha)(2t+u^2+t^2 v^2)-\alpha(2tuv+1+t^2) \right]\\\nonumber
&+{1+m\over 2}\ln \left(
f(ue^{\beta}) \right)+{1+m\over 2}\ln \left(t^c \,f(ve^{\beta})
              \right)\\\nonumber
&-{1+m\over 2}\ln \left({1+m\over 2} \right)-{1-m\over 2}\ln \left({1-m\over 2}
 \right)
\end{eqnarray}
with respect to the parameters $u,\ v,\ t$ and $\beta$.
The magnetization of the metastable states at energy $E$ is given by the
maximization of (\ref{eqnsm}) with respect to $m$:
\begin{equation}\label{mdee}
m ={f(u e^{\beta})-t^c\,f(v e^{\beta}) \over 
f(u e^{\beta})+t^c \,f(v e^{\beta})}
\end{equation}

In order to compute the Edwards entropy, we  solve the stationarity
conditions coming from the extremization with respect to $u,v,t,\beta$ 
of $S(u,v,t,\beta;E,m(E))$, with $m(E)$ replaced by the expression
(\ref{mdee}).  
The resulting value for $\beta$ is $\beta={\partial s_{Edw}(E) 
\over \partial E}$, which is the definition of the inverse
Edwards temperature.

\begin{figure}
\narrowtext
\epsfxsize=0.7\hsize
\epsfbox{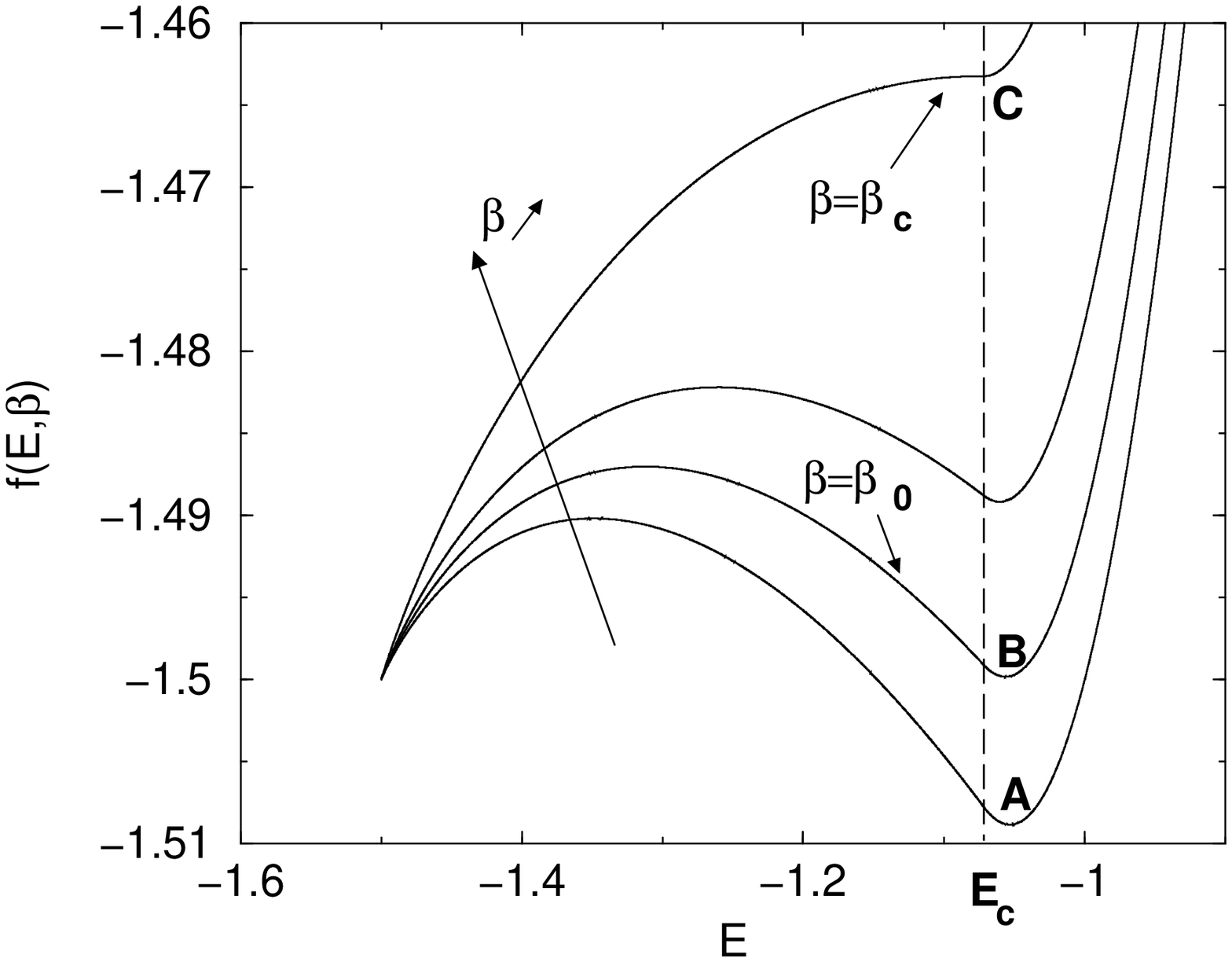}
\caption{}
\label{fig1}
\end{figure}

\begin{figure}
\narrowtext
\epsfxsize=0.7\hsize
\epsfbox{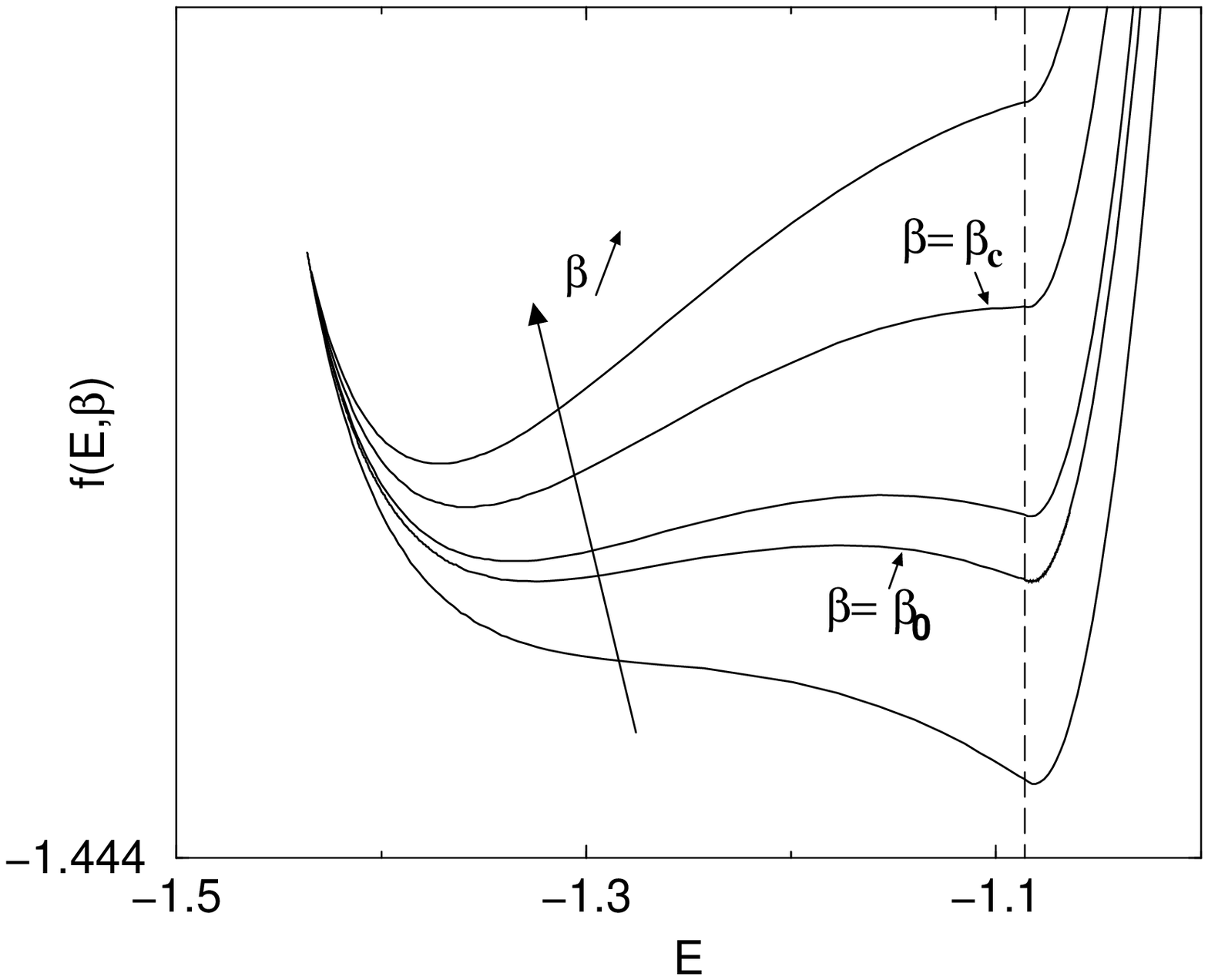}
\caption{}
\label{fig2}
\end{figure}

\begin{figure}
\narrowtext
\epsfxsize=0.7\hsize
\epsfbox{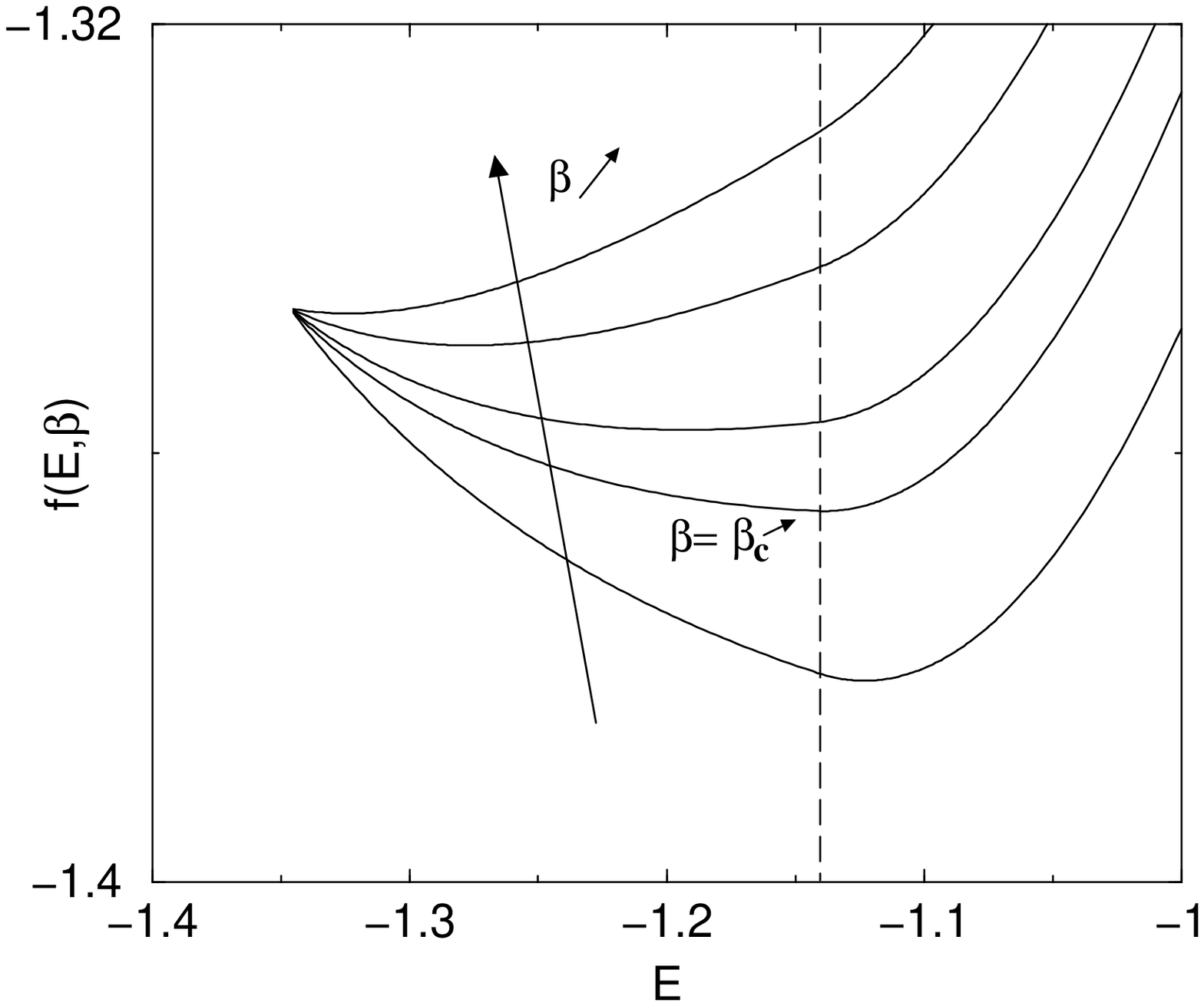}	
\caption{}
\label{fig3}
\end{figure}

The saddle point can be found numerically to compute 
$f(E,\beta)$. In the annealed approximation when $\alpha <
\alpha_c \simeq 0.06$ we find that we are in scenario (2), for
$\alpha > \alpha_c$ we are in scenario (1). 
Shown in Figs. (\ref{fig1}), (\ref{fig2}) and (\ref{fig3})
is the behavior of $f(E,\beta)$  for $\alpha = 0$,
$\alpha = 0.04$ and  $\alpha = 0.12$ respectively and one
sees that one has the behaviors (2b), (2a) and (1). In the range
$\beta_0 < \beta < \beta_c$ shown  in Figs. (\ref{fig1}) and (\ref{fig2})
the higher energy, paramagnetic, minimum is only metastable.

\begin{figure}
\narrowtext
\epsfxsize=0.7\hsize
\epsfbox{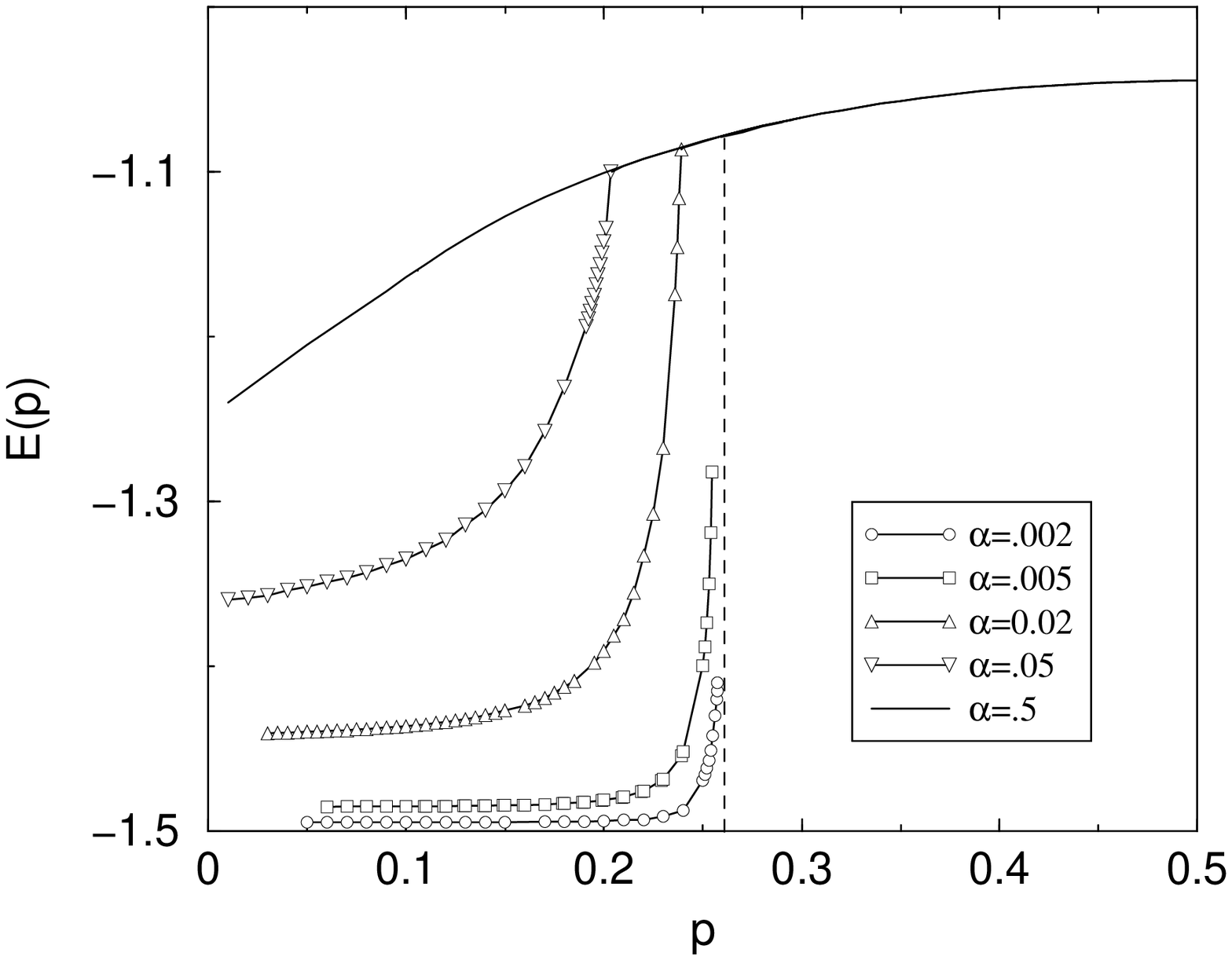}
\caption{}
\label{fig4}
\end{figure}

To test the above predictions, we have tapped systems with 
$\alpha=0.002, 0.005, 0.02, 0.05$ and recorded both the energy $E(p)$ and
the magnetization $m(p)$ in the quasi equilibrium regime. 
Systems were of size $N=10^6$ and we have
checked that this was large enough to consider a single sample instead of
several realizations of disorder. The curves for $E(p)$ are shown in
Fig. (\ref{fig4}). For $\alpha=0.005$ and 
$\alpha=0.002$, there is a first order phase transition, whereas for  
$\alpha=0.05$ and $\alpha=0.02$ this transition is a second order one, in
qualitative agreement with the previous discussion of the role of the
Edwards entropy. 
Of course, as the annealed entropy is not expected  to be exact a low
energy, the values of $\alpha_c$ and $E_c(\alpha)$
which come from this calculation are only approximate. 
Indeed, it is found in the tapping
simulations that $0.005<\alpha_c<0.02$, and $E(p_c(\alpha)^+,\alpha)$ differs
systematically with $E_c(\alpha)$ found from Eq. (\ref{eqnedw}).

\begin{figure}
\narrowtext
\epsfxsize=0.7\hsize
\epsfbox{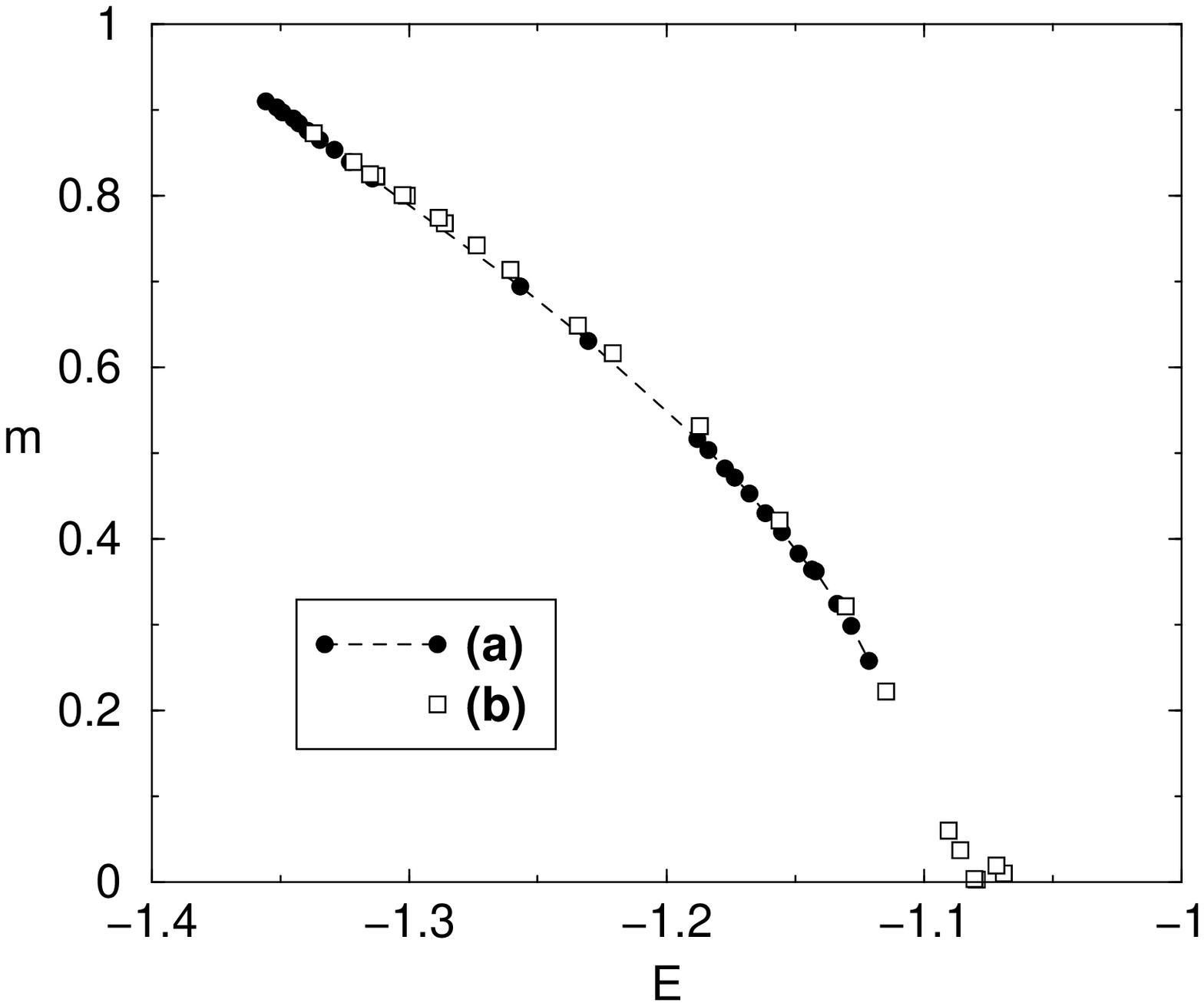}
\caption{}
\label{fig5}
\end{figure}

\begin{figure}
\narrowtext
\epsfxsize=0.7\hsize
\epsfbox{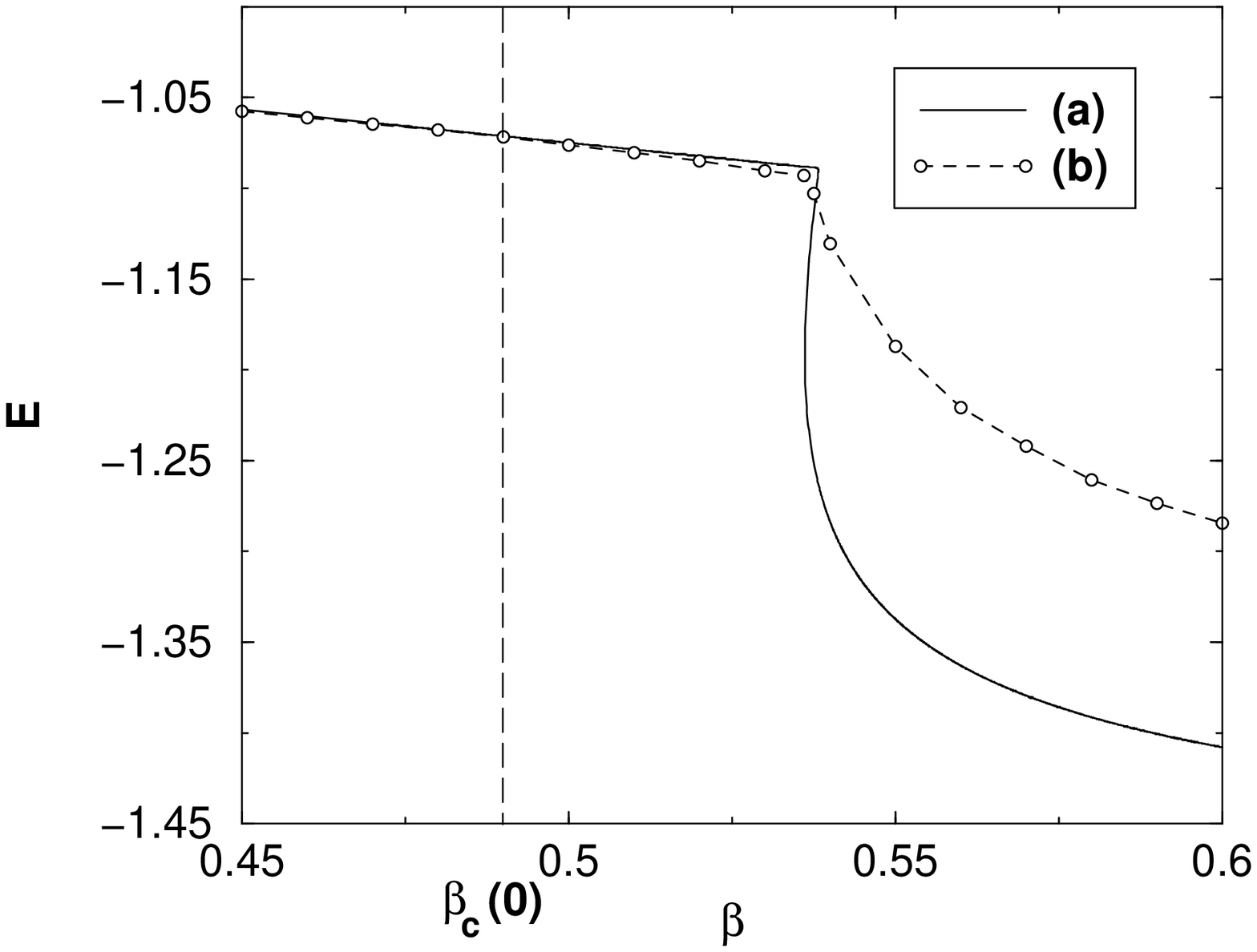}
\caption{}
\label{fig6}
\end{figure}

In order to test this scenario in a quantitative fashion,
we have computed the Edwards entropy as was done in
\cite{bar}. One introduces an auxiliary temperature $1/\beta_{aux}$, which
is a Lagrange multiplier to fix the number of spins which are not in the
same direction as their local field. Then, one fixes $\beta$ and does a
Monte-Carlo simulation using the Metropolis algorithm where a 
randomly chosen spin is flipped with probability 
$\mbox{min}\left( 1, e^{-\beta_{aux} \Delta {\cal H}_{aux}} \right)$, with
$\beta_{aux} {\cal H}_{aux}=\beta {\cal H}
+\beta_{aux} \sum_i \theta \left( s_i \sum_{j\neq i} n_{ij} J_{ij} s_j
\right)$. Starting from $\beta_{aux}=0$, we increased it very slowly until the
system was blocked in a metastable state. Repeated several times, this
allowed us to sample $E(\beta)$ for systems of $N=10^5$. 
At the end of the simulation, we have
recorded both energy and magnetization. The comparisons between $m(E(p))$ from
tapping simulations and $m(E(\beta))$ from Monte-Carlo simulations for 
$\alpha=0.05$ are shown in Fig.(\ref{fig5}), the agreement is excellent. 
In the high temperature region $\beta \in [0,\beta_c(0) ]$, for all 
$\alpha \in [0, 1/2]$, the value of $E(\beta)$ measured in the simulations
were in excellent agreement with the results of the annealed calculation,
indicating the latter is correct in this range. For $\beta_c(0)< \beta <
\beta_c(\alpha)$ (here we mean $\beta_c$ from the annealed approximation)
the annealed calculation is close but certainly deviates from the numerically
measured one (see Fig (\ref{fig6})). Moreover,  
the value of $E_c(\alpha)$ found in this Monte-Carlo procedure 
was much closer to $E(p_c(\alpha)^+)$ than that found in the 
annealed approximation. This means that this approximation is not
exact for $E$ when $m(E)\neq 0$ and explains the quantitative disagreement
between $E(p_c(\alpha)^-, \alpha)$ and $E^*(\alpha)$ obtained 
from (\ref{eqnsm}).

\begin{figure}
\narrowtext
\epsfxsize=0.7\hsize
\epsfbox{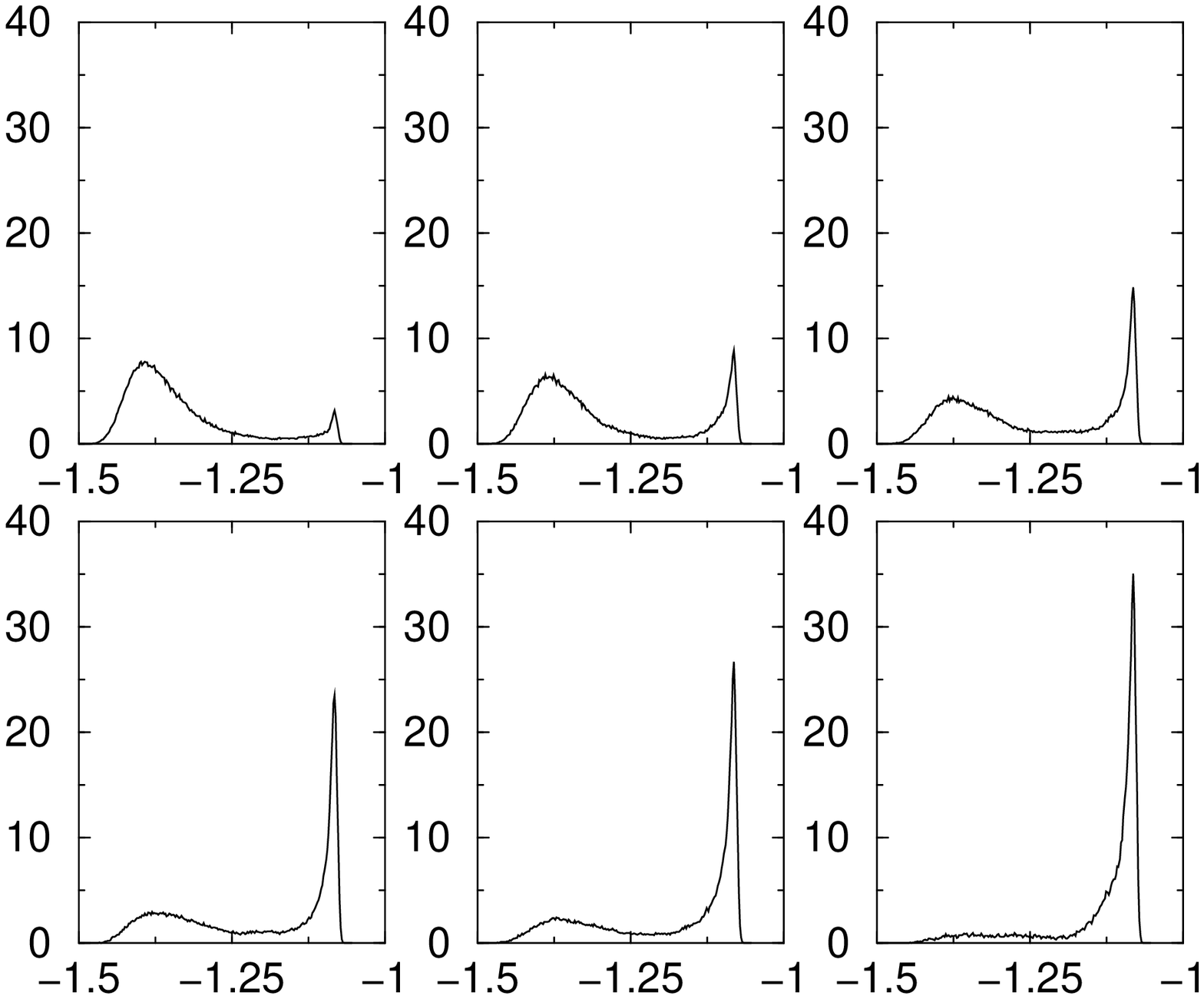}
\caption{}
\label{fig7}
\end{figure}

For $\alpha<\alpha_c$, the simulated annealing again predicts $E_c(\alpha)$
very close to  $E(p_c(\alpha)^+, \alpha)$ determined by the tapping
simulations. The behavior of the low energy magnetization for $E< E^*(\alpha)$
is also the same for the two simulations, although the region of energy 
close to $E^*(\alpha)$ is difficult to sample with the simulated annealing.
Fig.(\ref{fig7}) is the  histogram of 
the energy obtained during the tapping dynamics in the stationary regime 
for $\alpha=0.002$ for different $p$ near $p_c$, for a relatively small
system size. There are clearly two peaks separated by a gap and the 
spread towards negative energy 
of the one of highest energy becomes large when $p$ 
approaches $p_c$, thus indicating the approach of the spinodal point.
It is interesting to compare Fig. (\ref{fig2}) qualitatively  with 
Fig.(\ref{fig7}), the comparison is quite striking. 
For $\beta $ slightly smaller than
$\beta_c$ we see that the form of the high energy peak, with a small
spread to the right and a large one to the left is predicted by the 
geometry of the first minimum of $f(E,\beta)$ in Fig. (\ref{fig2}). In
addition for $\beta $ slightly smaller than $\beta_c$ we see that the 
ferromagnetic minimum is close to its spinodal point, thus
explaining the large, two sided in this case, spread in the histogram
in this region. 

In conclusion we have examined a dynamics driven by an extensive 
external perturbation on a family of systems having an extensive 
Edwards entropy of blocked states. The existence of first and
second order ferromagnetic transitions are qualitatively explained
by invoking the Edwards measure calculated within an annealed approximation.
The quantitative predictions using this measure in Monte-Carlo simulations
are also in excellent agreement with tapping simulations on the system. 
We have found similarly promising results in models of higher connectivity and
in the infinitely connected SK spin glass \cite{wip}. This work suggests
the possibility of using the Edwards measure to explain phase transitions
in realistic granular systems, the main obstacle is the technical difficulty
in the computation of the Edwards measure.

\end{multicols}
{\bf Figure captions}

Fig. 1. Free energy $f(E,\beta)$ versus $E$ for different values of $\beta$
	with $\alpha=0$. Between A and B, the local minimum has the lowest
	free energy, between B and C, it is only metastable and above C it
	is no longer a minimum. 
  
Fig. 2. Free energy $f(E,\beta)$ versus $E$ for different values of $\beta$
	and $\alpha=0.04 < \alpha_c$. One sees that when 
	$\beta$ increases the minimum of high energy
	starting as a global minimum becomes metastable and then disappears
	at the transition.   

Fig. 3.	Free energy $f(E,\beta)$ versus $E$ for different values of $\beta$
	and $\alpha=0.12 > \alpha_c$. Here there is only one local minimum 
	which moves continuously with $\beta$. 
	The vertical dashed lines indicates the energy $E$ where
	$f(E,\beta)$ is non analytic.	

Fig. 4. Energy per spin $E(p)$ in the steady state regime versus $p$ for
	$\alpha=0.002, 0.005, 0.02, 0.05, 0.5$ and $N=10^6$ spins.

Fig. 5. Comparison of the measurements of $m(E)$ in the tapping simulations
	and Monte-Carlo simulations. In the former (a) $m(p)$ is plotted
	versus $E(p)$ and in the latter (b) $m(\beta)$ is plotted
	versus $E(\beta)$.
 	
Fig. 6. Comparison of $E(\beta)$ for $\alpha=0.05$ obtained from the
	annealed calculation (a) and the Monte-Carlo simulation (b). At the
	right of the vertical dashed line, the annealed calculation is not
	exact any more. It predicts a first order phase transition, whereas
	the Monte Carlo predicts a second order one, with $E_c$ in very good
	agreement with that of the tapping at $p=0.05$. 

Fig. 7. Histogram of the energy during the tapping simulations in the steady
	state regime for $\alpha=0.002$ and $N=25000$ spins during $10^5$
	taps and for $p=0.2584, 0.2585, 0.2586, 0.2587, 0.2588, 0.2589$
        (reading left to right from the top).  
 
\begin{multicols}{2} 
  
\end{multicols}

\end{document}